\documentclass[twocolumn,showpacs,nofootinbib,aps,prd]{revtex4}

\usepackage{graphicx,url,color}
\usepackage{amsmath,amsthm,amssymb}

\renewcommand{\vec}[1]{\boldsymbol{#1}}

\definecolor{Black}{named}{Black}
\definecolor{Blue}{named}{Blue}
\definecolor{Red}{named}{Red}

\newcommand{\be}{\begin{equation}}
\newcommand{\ee}{\end{equation}}
\newcommand{\ba}{\begin{eqnarray}}
\newcommand{\ea}{\end{eqnarray}}

\def\lsim{\raise0.3ex\hbox{$\;<$\kern-0.75em\raise-1.1ex\hbox{$\sim\;$}}}
\def\gsim{\raise0.3ex\hbox{$\;>$\kern-0.75em\raise-1.1ex\hbox{$\sim\;$}}}

\def\theta{\vartheta}

\def\F{{\cal F}}

\def\aap{Astron.\ Astrophys.\ }
\def\mnrs{Mon.\ Not.\ Roy.\ Soc.\ }

\begin{document}

\title{Deriving the cosmic ray spectrum from gamma-ray observations}

\author{M.~Kachelrie\ss$^{1}$}
\author{S.~Ostapchenko$^{1,2}$}
\affiliation{$^1$Institutt for fysikk, NTNU, Trondheim, Norway}
\affiliation{$^2$D.~V.~Skobeltsyn Institute of Nuclear Physics, Moscow
State University, Moscow, Russia}

\begin{abstract}
A fundamental problem of cosmic ray (CR) physics is the determination of the 
average properties of Galactic CRs outside the Solar system. Starting from 
COS-B data in the 1980's, gamma-ray observations of molecular clouds in the 
Gould Belt above the Galactic plane have been used to deduce the Galactic CR 
energy spectrum. We reconsider this problem in view of 
the improved precision of observational data which in turn require a more 
precise treatment of photon production in proton-proton scatterings. 
We show that the spectral shape $dN/dp\propto p^{-2.85}$ of CR protons as 
determined by the PAMELA collaboration in the energy range 
$80\,{\rm GeV}<p{\rm c}<230$\,GeV is consistent with 
the photon spectra from molecular clouds observed with Fermi-LAT down to 
photon energies $E\sim 1$--2\,GeV. Adding a break of the CR flux at 3\,GeV,
caused by a corresponding change of the diffusion coefficient, improves
further the agreement in the energy range 0.2--3\,GeV.
\end{abstract}

\date{May 24, 2012}


\maketitle

\section{Introduction}

The propagation of cosmic ray (CR) protons and nuclei with energies 
$E/Z\lsim 10^{18}$\,eV in the turbulent component of the Galactic 
magnetic field  resembles a random walk and can be described in general  
by the diffusion approximation~\cite{diff}. Therefore the Galactic disk
should be filled with a well-mixed ``sea'' of CRs   whose  properties
are summarized by the differential diffuse intensity $I(E)$ or the differential
number density $n(E)=4\pi\, I(E)/v$. Excluding the regions close to recent
CR sources, the gradient $\vec\nabla \ln[n(E)]$ induced by the small current
of CRs diffusing out of the disc and its extended CR halo is
small.

Most of our knowledge about Galactic CRs is obtained via 
observations in our local environment. Despite of the diffusive 
propagation of CRs, the CR intensity deduced locally may differ 
from the one averaged over the Galactic disc: 
At low energies, the influence of the Solar wind on
measurements of the CR energy spectrum and the total
CR energy density has to be corrected based on current understanding 
of the heliospheric modulation  and direct CR measurements at different 
heliospheric distances and at different modulation levels.
Clearly, such a correction is model-dependent and can introduce 
uncertainties.
Moreover, the Sun is close to a region with increased star-formation
and thus supernova rate. Other reasons for local deviations include 
stochastic re-acceleration in the local interstellar turbulence
or local sources as old supernova shocks, winds and flares of massive stars.
Therefore, the CR density close to the Solar
system may deviate from the average Galactic one.

A way to obtain independent information on the average ``sea'' Galactic CRs 
is the observation of suitable molecular clouds far from CR accelerating 
regions~\cite{review}: These clouds serve as a target for CRs producing  
gamma-rays mainly through decays of neutral pions created as secondaries 
in CR-gas collisions. Suitable clouds should be located  away from the 
Galactic plane in order to test the ``sea'' CR spectrum, excluding the
directions towards the inner and outer Galaxy. 
Assuming that gamma-ray production in hadronic interactions is sufficiently
well-understood, the observed gamma-ray flux $\F_\gamma(E)$ from these clouds 
can be inverted to obtain the differential CR number density $n(E)$.

Previous works used observations of molecular clouds in the 
Gould Belt, in particular of  Orion A and B, performed first
by COS-B~\cite{1980A&A....91L...3C}, then
EGRET~\cite{1995ApJ...441..270D,1999ApJ...520..196D}, 
and most recently Fermi-LAT
to derive the spectral shape of
Galactic CRs. During this period, the quality of experimental data has been
hugely improved: On the observational side, the data from Fermi-LAT have 
a much reduced error compared to its predecessor EGRET and extend
now up to photon energies $E_\gamma\sim 100$\,GeV, corresponding to
typical energies of CR primaries $E\sim 1$\,TeV. Moreover, the 	
PAMELA collaboration determined the slope $\beta_{\rm CR}$ of the
CR spectrum $dN/dp\propto p^{-\beta_{\rm CR}}$ 
with an accuracy of $\Delta\beta_{\rm CR}=\pm 0.05$ in the energy range
$80\,{\rm GeV}<p{\rm c}<230$\,GeV~\cite{pamela}.
Thus the prediction of the secondary photon spectrum requires either
similar precise photon fragmentation functions, or at least an estimate of
their error. Finally, there are new results  on photon yields from
HERA~\cite{HERA} as well as from LHC~\cite{LHCf} on the accelerator side, 
restricting theoretical models for photon fragmentation functions. 

In view of the improved precision of the experimental data, 
we reconsider this problem, paying special attention to the treatment of 
photon production in proton-proton scatterings. We find that
several commonly used parametrisations for the 
photon fragmentation function as the ones of Refs.~\cite{kamae07,kelner}
deviate substantially from  experimental data at high energies.
These differences diminuish considering the photon yield produced by
CRs with a power-law momentum distribution. In this case,
we find a relatively good agreement concerning the shape
of the photon spectra, while the absolute photon yield differs by
$\sim 20$\%.
As our main result, we show that the spectral shape $dN/dp\propto p^{-2.85}$ 
of CR protons as determined 
by PAMELA in the energy range $80\,{\rm GeV}<p{\rm c}<230$ GeV~\cite{pamela}
is consistent with 
the photon spectra from molecular clouds observed by Fermi-LAT down to 
energies $E\sim 1$--2\,GeV. The agreement is further improved, if the
CR spectrum exhibits a break around 3\,GeV, as suggested by radio
data~\cite{low2}.

This work is structured as follows: We compare first in 
Sec.~\ref{sec:Model-approaches} several models used for the calculation 
of photon production in hadronic collisions to experimental data. 
We conclude that a combination of  the parametrization of 
Ref.~\cite{kamae07} for nondiffractive processes below $E_{{\rm thr}}=50$\,GeV
with the QGSJET-II model~\cite{ost11} at higher energies
 gives a satisfactory description of experimental
data. Then we calculate in Sec.~\ref{sec:gamma-ray spectra} the photon
spectra expected from molecular clouds for a given CR flux.
In the appendix, we describe the use of the photon and antiproton 
fragmentation functions employed by us which are available from
\url{http://sourceforge.net/projects/ppfrag}.

\section{Models for photon production}
\label{sec:Model-approaches}

High-energy photons can be produced both by CR protons and electrons. In the 
latter case, inverse Compton scattering on photons mainly from the cosmic 
microwave 
background and bremsstrahlung are potentially contributing processes.
In particular, bremsstrahlung was discussed as an important contribution
to the total observed gamma-ray spectrum from molecular clouds at energies 
below 100\,MeV~\cite{1999ApJ...520..196D}. 
In this work, we restrict ourselves however 
to the energy range $E_\gamma>200$\,MeV observed by Fermi-LAT where
bremsstrahlung can be neglected. For the density of molecular clouds,
also inverse Compton scattering gives a negligible contribution
relative to photon production in CR-gas collisions. 

Photon production in 
hadronic collisions results mainly from decays of neutral pions produced
as secondaries. At sufficiently high energies, an additional though much 
smaller contribution comes from $\eta$ decays, while direct photon 
production is strongly suppressed and negligible for astrophysical 
applications%
\footnote{In contrast, the results of Ref.~\cite{huang} seem to indicate
a significant contribution from direct photon production. This surprising
result is explained by the simple fact that the  authors of  Ref.~\cite{huang}
refer misleadingly to photons from  $\eta$ decays as direct photons.}.

The photon yield in hadronic collisions can be calculated using
either numerical parametrisations or Monte Carlo simulations. The former are 
typically based on theoretically motivated or empirical scaling laws 
fitted to accelerator data. In general, they are  well-suited for collisions 
  at relatively low energies  $E\lsim 50$\,GeV.   
In contrast, Monte Carlo simulations  
are developed mainly for high-energy collisions and are based on a 
combination of non-perturbative models and perturbative QCD.
They treat both soft
interactions and  the hadronization of partons produced in (semi-) hard
scattering processes.

In the next subsection,  we compare 
the results derived using the ``Kamae'' parametrisation for the 
photon fragmentation function given in  Ref.~\cite{kamae07} to those
obtained from the Monte Carlo simulations
   with the QGSJET-II-04~\cite{ost11}
and SIBYLL 2.1 \cite{sibyll} models, the latter having been used as the 
basis for the parametrization of the photon fragmentation function of
Ref.~\cite{kelner}.

\begin{table}
\begin{tabular}{lccccc}
\hline 
 & 10\,GeV & 100\,GeV & 1\,TeV & 10\,TeV & 100\,TeV\tabularnewline
\hline 
QGSJET-II-04 & 0.62 & 0.64 & 0.64 & 0.70 & 0.79\tabularnewline
SIBYLL 2.1 & - & 0.71 & 0.84 & 0.93 & 1.1\tabularnewline
Model of Ref.~\cite{kamae07} & 0.57 & 0.78 & 0.73 & 0.84 & 1.1\tabularnewline
ND model of~\cite{kamae07}& 0.48 & 0.53 & 0.48 & 0.49 & 0.69\tabularnewline
\hline
\end{tabular}
\caption{Predictions for the spectral-weighted moment $Z_{\gamma}(E_{0})$
(in mb) for photon production in $pp$ collisions at different   laboratory
energies $E_0$:
 Comparison
of QGSJET-II-04, SIBYLL 2.1, the parametrization of Ref.~\cite{kamae07},  and 
the nondiffractive (ND) model of Ref.~\cite{kamae07}; 
all for $\beta_{{\rm CR}}=-2.85$.}
\label{Flo:zfac}
\end{table}

\subsection{Comparison of the Kamae parametrisation and QGSJET-II}

We start by recalling some basic analytical formula before we discuss our 
numerical results for the various quantities characterizing photon production 
in hadronic collisions. Assuming a power-law cosmic ray spectrum, 
$dN_{{\rm CR}}/dE=N_{0}\, E^{-\beta_{{\rm CR}}}$,
the resulting $\gamma$-ray flux may be written as
\begin{eqnarray}
\frac{E_{\gamma}^{2}\, dN_{\gamma}}{dE_{\gamma}} & \propto & E_{\gamma}^{2}\,\int_{E_{\gamma}}^{E_{\max}}\! dE'\;\frac{dN_{{\rm CR}}}{dE'}\,\frac{d\sigma^{pp\rightarrow\gamma}(E',E_{\gamma})}{dE_{\gamma}}
\nonumber \\
 & \propto & E_{\gamma}^{2-\beta_{{\rm CR}}}\int_{0}^{1}\! dx_{E}\;
 \frac{x_{E}^{\beta_{{\rm CR}}-1}\, 
 d\sigma^{pp\rightarrow\gamma}(E_{\gamma}/x_{E},x_{E})}{dx_{E}}
\nonumber \\
& \equiv & E_{\gamma}^{2-\beta_{{\rm CR}}}\,\tilde{Z}_{\gamma}(E_{\gamma})\,,
 \label{eq:gamma-spec}
 \end{eqnarray}
where $x_{E}=E_{\gamma}/E'$ and we have assumed $E_{\max}\gg E_{\gamma}$.
Thus, any difference in the spectral shape between the parent CRs
and the produced photons is introduced by the violation of Feynman scaling, 
i.e.\ by  the energy dependence of the spectral moment 
$\tilde{Z}_{\gamma}$. Such a dependence emerges because of 
i) the slow energy rise of the inelastic $pp$ cross section 
$\sigma_{pp}^{{\rm inel}}(E)$,
ii) the relatively fast increase of the central rapidity plateau of
secondary particles 
$1/\sigma_{pp}^{{\rm inel}}\, d\sigma_{pp}^{\pi^{0}(\eta)}/dy |_{y=0}$,
and iii) the slow ``softening'' of the forward spectra of secondary
mesons. While the first two effects result in an energy rise of 
$\tilde{Z}_{\gamma}$, the third one works in the opposite direction.

For our purposes, it is more convenient to consider moments $Z_{\gamma}(E_{0})$
defined for a given energy $E_0$ of the CR proton in the $pp$ collision,
\begin{equation}
 Z_{\gamma}(E_{0}) = \int_{0}^{1}\! dx_{E}\;\frac{x_{E}^{\beta_{{\rm CR}}-1}\, d\sigma^{pp\rightarrow\gamma}(E_{0},x_{E})}{dx_{E}}\,.
\label{eq:moment}
\end{equation}
In Table \ref{Flo:zfac},
we illustrate the energy dependence of $Z_{\gamma}(E_{0})$, using
$\beta_{{\rm CR}}=-2.85$ as reported by the Pamela Collaboration
in the range $50\:{\rm GeV}<E_{0}<200\:{\rm GeV}$ \cite{pamela}, comparing
the predictions of the QGSJET-II-04  \cite{ost11} and SIBYLL~2.1 \cite{sibyll} 
Monte Carlo generators
to the results obtained using the parametrization 
of $d\sigma^{pp\rightarrow\gamma}(E_{0},E_{\gamma})/dE_{\gamma}$
from Ref.~\cite{kamae07}. Clearly, the considered models predict a quite
different behavior of $Z_{\gamma}(E_{0})$ in the energy range of
interest: While in the case of QGSJET-II the spectral moment is approximately
energy-independent up to $E_0\sim 1$ TeV,  $Z_{\gamma}(E_{0})$ has a relatively
steep energy rise in the other two models.

\begin{figure*}[tbh]
\includegraphics[width=\textwidth]{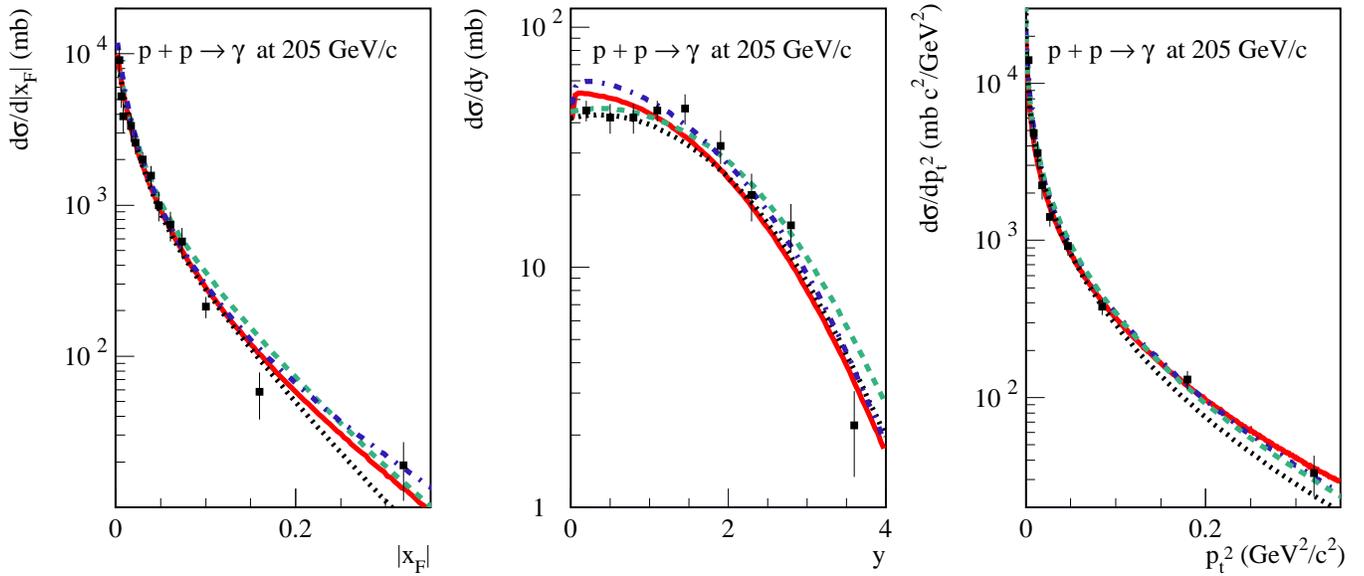}
\caption{Photon production cross sections in the center of mass (c.m.) frame
for proton-proton collisions at 205 GeV/c:   Feynman $x$ spectrum (left),
 rapidity distribution (middle), and
 transverse momentum distribution (right). Calculations with QGSJET-II-04
(red solid),  SIBYLL~2.1 (blue dot-dashed), parametrization of Ref.~\cite{kamae07} (green dashed), and ND part (black dotted line) of the latter
 are shown together with experimental data from Ref.~\cite{jaeger}. 
\label{fig:200}}
\end{figure*}

\begin{figure*}[tbh]
\includegraphics[width=0.8\textwidth]{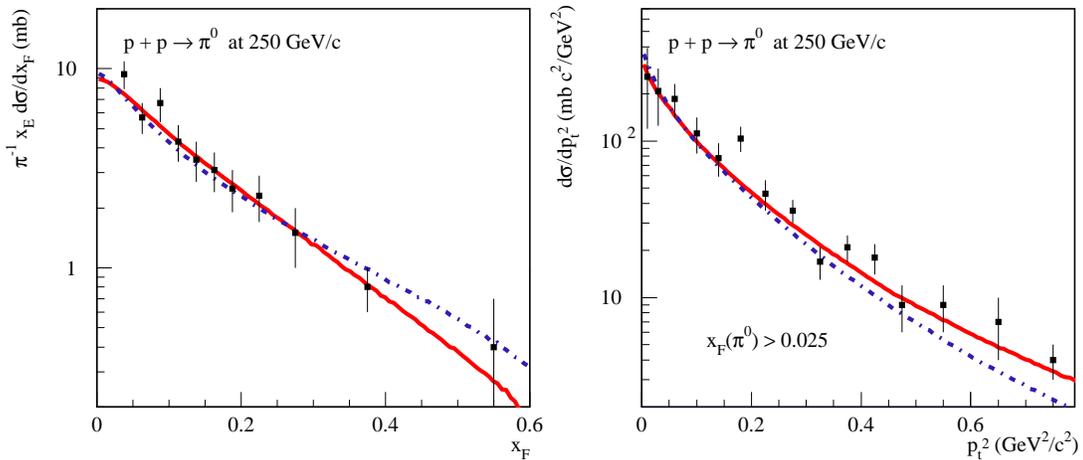}
\caption{Feynman $x$ spectrum (left) and  transverse momentum distribution 
(right) of neutral pions in the c.m.\ frame for $pp$ collisions at 250 GeV/c
 as calculated using
QGSJET-II-04 (red solid) and SIBYLL~2.1 (blue dot-dashed)
compared to the data from Ref.~\cite{ajinenko}.
\label{fig:250}}
\end{figure*}%
\begin{figure*}[tbh]
\includegraphics[width=0.8\textwidth]{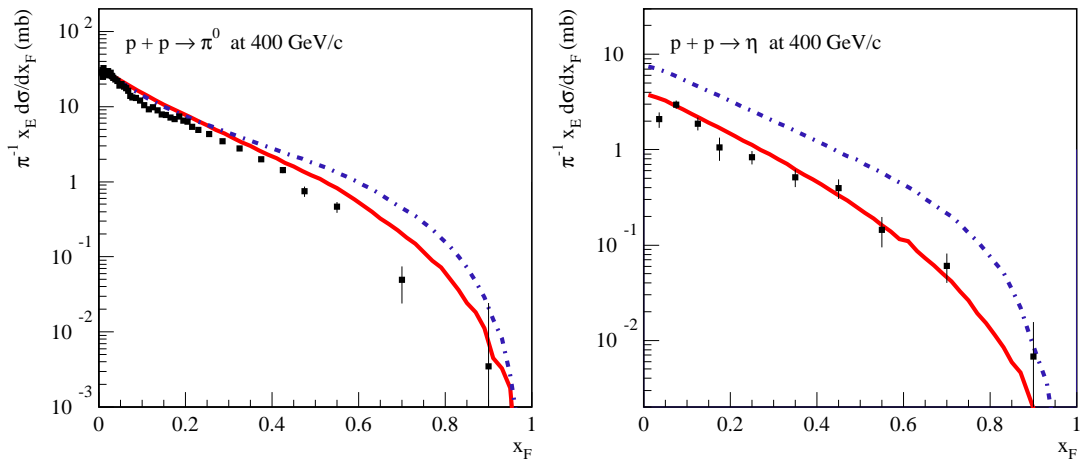}
\caption{Feynman $x$ spectra for $\pi^{0}$ (left) and $\eta$
(right) production in the c.m.\ frame for $pp$ collisions at 400 GeV/c
 as calculated using QGSJET-II-04 (solid) and SIBYLL~2.1 (dot-dashed)
compared to the data from Ref.~\cite{ehs400}.\label{fig:400}}
\end{figure*}

\begin{figure*}[tbh]
\includegraphics[width=0.9\textwidth]{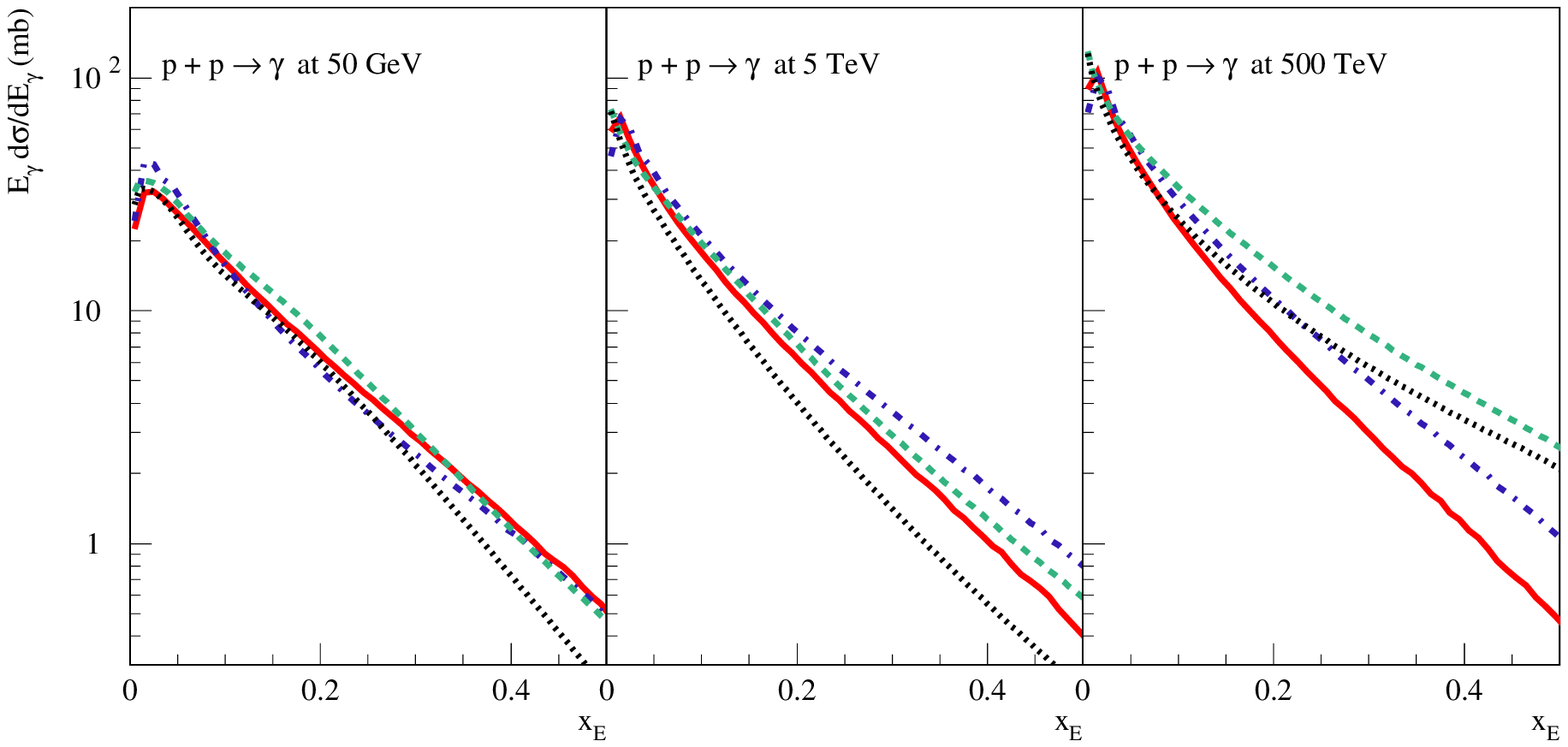}
\caption{Calculated energy distributions of photons in the laboratory frame 
for $pp$ collisions at 50 GeV, 5 TeV, and 500 TeV;
abbreviations for the lines are the same as in Fig.\  \ref{fig:200}.
\label{fig:gam-lab}}
\end{figure*}

In order to decide which of these approaches provides a better
description of photon production, we compare them next to 
data from accelerator experiments. We start by considering the respective
results for  photon spectra in proton-proton collisions at 205 GeV/c
laboratory momentum in Fig.~\ref{fig:200}  and 
for   spectra of neutral pions and etas  in $pp$  collisions at
 $p_{\rm lab}=250$ and 400 GeV/c in  Figs.~\ref{fig:250} and \ref{fig:400}.
In the case of QGSJET-II, the comparison demonstrates a good overall agreement 
between the predictions and the experimental data. 
The Feynman $x_{\rm F}$ spectra of photons and neutral pions  
obtained with SIBYLL~2.1 are very similar to the 
QGSJET-II case at small   $x_{\rm F}$, but become substantially harder
with increasing energy for $x_{\rm F}\gsim 0.2$, see 
Figs.~\ref{fig:200}--\ref{fig:gam-lab}.

As the accuracy of the fixed-target data considered
does not allow one to discriminate between these two trends, a valuable
benchmark is provided by recent HERA measurements of photon production in the
proton fragmentation region for $p\gamma ^*$ interaction at $\sqrt s=319$ GeV
\cite{HERA}: The observed forward energy spectra of gammas
appear to be well-described by QGSJET-II, while being substantially softer
than  SIBYLL predictions. 

At first sight, this conclusion seems to be in a contradiction
with the results of spectrometer studies of   $pp$ collisions
at $\sqrt s=7$ TeV by the LHCf collaboration \cite{LHCf}:
The measured very forward photon spectra proved to be significantly harder
than predicted by  QGSJET-II. However, the preliminary results on the 
$\pi ^0$ production at $\sqrt s=7$ TeV by the same collaboration
indicate that the latter discrepancy is likely due to a somewhat softer than
observed transverse momentum distribution of neutral pions in  
QGSJET-II~\cite{lhcf-pi}\footnote{Fixed angle spectrometer measurements 
of forward particle
spectra are very sensitive both to the respective Feynman $x$ and $p_t$
distributions, the two variables being related to each other in the c.m.\ frame
as $p_t=\theta _{\rm obs}\: x_{\rm F}\:\sqrt s/2$.}.
 The latter conjecture
 is supported also by the HERA data of Ref.~\cite{HERA} which demonstrate that 
the $p_t$-spectra of forward photons are better described by  SIBYLL~2.1
 than by QGSJET-II. We conclude  that QGSJET-II agrees well with the 
experimental data on the energy spectra of photons. The predicted
$p_t$ distributions are somewhat too soft, but are anyway
 irrelevant for our problem.

\begin{figure*}[tbh]
\includegraphics[width=\textwidth]{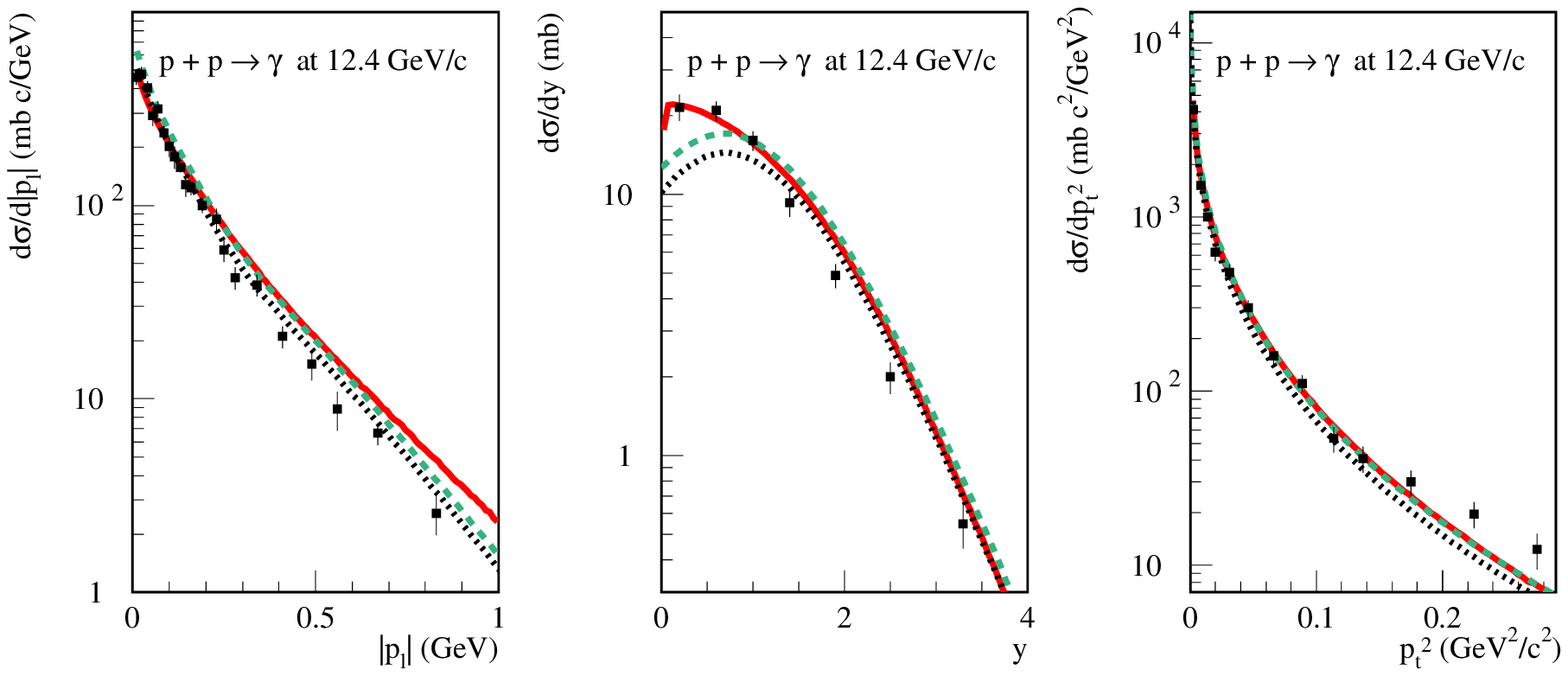}
\caption{Longitudinal  momentum spectrum  (left), rapidity 
distribution (middle),
  and transverse momentum distribution (right) of photons in the c.m.\ frame
  for $pp$
collisions at 12.4 GeV/c as calculated using QGSJET-II-04 (solid lines)
compared to experimental data from Ref.~\cite{jaeger12}. Dashed and
dot-dashed lines - parametrizations of Ref.~\cite{kamae07} for photon production
in inelastic and nondiffractive $pp$ collisions, respectively.
\label{fig:12}}
\end{figure*}

\begin{figure*}[t]
\includegraphics[width=0.8\textwidth]{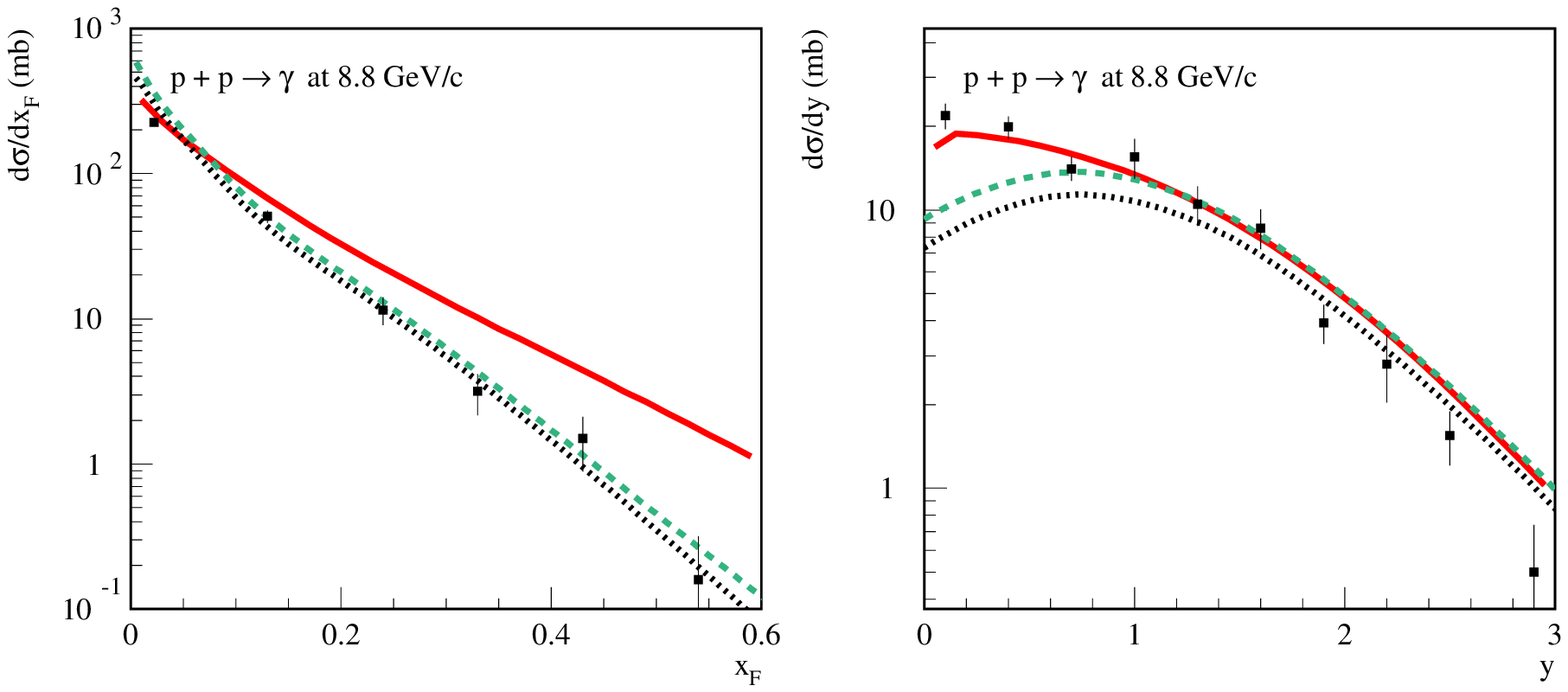}
\caption{Calculated Feynman $x$ (left) and rapidity (right) spectra of photons
in the c.m.\ frame for $pp$ collisions at 8.8 GeV/c   compared
to experimental data from Ref.~\cite{booth}; abbreviations for the lines
are the same as in Fig.~\ref{fig:12}.\label{fig:8}}
\end{figure*}

Though the considered Monte Carlo simulations are not designed 
to treat hadronic collisions for $E_0\lsim 50$ GeV, the QGSJET-II results
appear to be relatively reasonable down to $\sim 12$\,GeV/c laboratory momentum,
as demonstrated in Fig.~\ref{fig:12}.
However, extrapolating the model to even smaller energies is meaningless,
because the relevant physical processes, like resonance production and 
secondary Reggeon exchanges, are
not included. Indeed, as it can be clearly seen in Fig.~\ref{fig:8},
at 8.8\,GeV/c laboratory momentum QGSJET-II tends to predict a harder
photon spectrum than observed experimentally.

In contrast, the situation with the model of Ref.~\cite{kamae07} appears
to be  quite 
different: This parametrisation describes the experimental data quite well 
up to $E_{0}\sim10$\,GeV, cf.~Figs.~\ref{fig:12} and \ref{fig:8}. 
Figures~\ref{fig:200} and  \ref{fig:gam-lab} show that the  photon spectra 
predicted by the Kamae parametrization become at higher energies much harder 
than those of QGSJET-II which, as we have argued above, agree with HERA
data. This explains the larger values for $Z_{\gamma}(E_{0})$ obtained
in that case in the energy range 
$E_{0}\gtrsim100$\,GeV, see Table 1. The discrepancy between the parametrization
of Ref.~\cite{kamae07} and the data has its origin in the somewhat 
oversimplified treatment of diffractive particle production in the
underlying model~\cite{kamae05}
which utilises 
a cluster-like hadronization procedure: The energy of 
the diffractive state is distributed more or less uniformly between the 
pions produced, neglecting the leading baryon effect. 
In reality, a large
part of high energy diffraction corresponds to the creation of high mass
states which are described by the Pomeron contribution and correspond
to multiperipheral kinematics of particle production\footnote{Even more 
complicated diffractive final states are produced at much higher 
energies~\cite{ost11,ost10}.}~\cite{kai79}.
As a consequence,  forward spectra of neutral pions and etas in
high mass diffractive processes resemble the ones of the usual 
nondiffractive collisions.

The complementarity of the two models  motivated us to combine the 
QGSJET-II description with
the parametrisation of Ref.~\cite{kamae07}: While we use
QGSJET-II for photon production at relatively high energies
 $E_{0}>E_{{\rm thr}}$,
with $10\lesssim E_{{\rm thr}}\lesssim 50$~GeV, at lower energies
we apply the parametrization of Ref.~\cite{kamae07} restricted to
{\em nondiffractive\/} processes.

\section{Gamma-ray spectra\label{sec:gamma-ray spectra}}

\begin{figure}[t]
\includegraphics[width=0.45\textwidth]{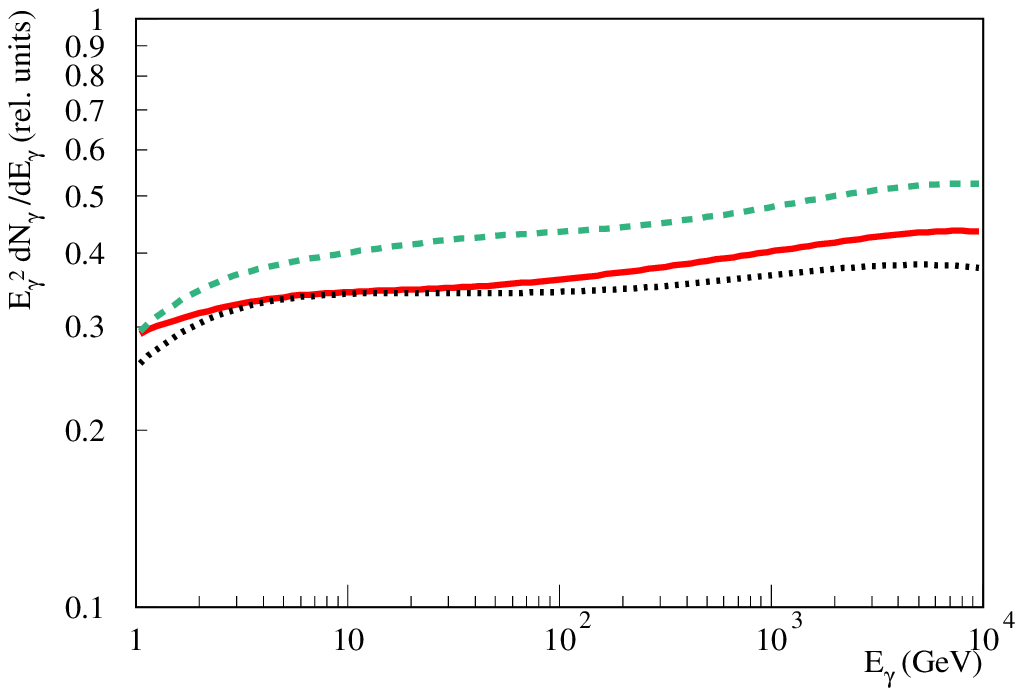}
\caption{Gamma-ray flux $E^2\F(E)$ produced by
cosmic ray protons with a spectrum   $dN/dp\propto p^{-2}$
as calculated using  QGSJET-II (solid red line),  
the Kamae parametrization \cite{kamae07} (dashed green line),  
or the nondiffractive part of the latter (dotted black line).
\label{fig:gamma-2}}
\end{figure}

We explore now the consequences of our new photon fragmentation function.
Let us remind first that diffusive shock acceleration predicts a power-law
in momentum, $dN/dp\propto p^{-\beta_{\rm CR}}$, cf. Ref.~\cite{bell}, 
 while the interstellar propagation is also rigidity-dependent. It
is, therefore, 
natural to use a power-law in rigidity to fit CR data. Indeed, PAMELA
p and He spectra and their ratio can be well-described with a power-law in
rigidity down to $\sim 20$\,GV,
below which the interstellar spectrum is significantly modified by the
heliospheric modulation.

In Fig.~\ref{fig:gamma-2}, we compare the gamma-ray spectrum produced by
cosmic rays with $dN/dp\propto p^{-2}$ using QGSJET-II,
the parametrization of Ref.~\cite{kamae07}, or using only the
 nondiffractive part of the latter. The photon 
 spectrum obtained using the Kamae parametrization rises much quicker with
 energy in the 1--10\,GeV range than the one obtained with  QGSJET-II,
 as a result how the diffraction is modeled in Ref.~\cite{kamae05}, 
 while at higher energies the obtained spectra have similar shapes,
 with $\sim 20$\% difference in the normalization. 
 On the other  hand, the results
 obtained using the nondiffractive part of the Kamae parametrization agree 
 well with  QGSJET-II  up to few tens of GeV -- compatible with 
 the similarity of the photon spectra for $pp$ collisons in the two models
 in the energy range $E_p\sim 10- 200$ GeV
  seen\footnote{The difference at $1- 3$ GeV is caused by the
 extrapolation of the QGSJET-II model outside its working range: Below 
 10\,GeV lab. energy,  it predicts   too hard photon spectra, 
 as shown in  Fig.~\ref{fig:8}.} in Figs.~\ref{fig:200}
 and
 \ref{fig:gam-lab}. Therefore, we find it natural to match 
 in the following
  the results of the QGSJET-II model for photon spectra
   to the parametrization
of Ref.~\cite{kamae07} for nondiffractive processes at $E_{{\rm thr}}=50$\,GeV.

\begin{figure}[t]
\includegraphics[width=8.5cm,height=7cm]{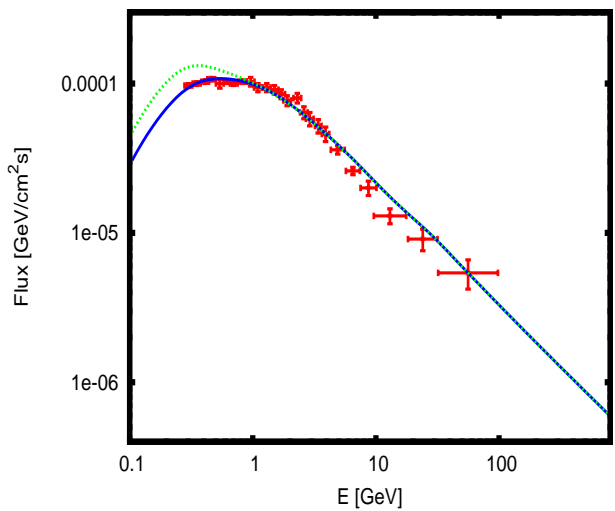}
\caption{Gamma-ray flux $E^2\F(E)$ as calculated using the combination
of   QGSJET-II  and the ND model of Ref.~\cite{kamae07}
($E_{\rm thr}=50$ GeV) for a single power-law CR flux with
  the slope $\beta_{{\rm CR}}=-2.85$  (dashed green line) and for a broken
  power-law with $\beta_{{\rm CR}}=-2.85$ for $p_{\rm CR}>3$ GeV/c and
  $\beta_{{\rm CR}}=-1$ for $p_{\rm CR}<3$ GeV/c (solid blue line).
The gamma-ray spectrum derived in Ref.~\cite{NST} from Fermi-LAT data
is shown as points with error-bars.
\label{fig:gamma-b1}}
\end{figure}

\begin{figure}[t]
\includegraphics[width=8.5cm,height=7cm]{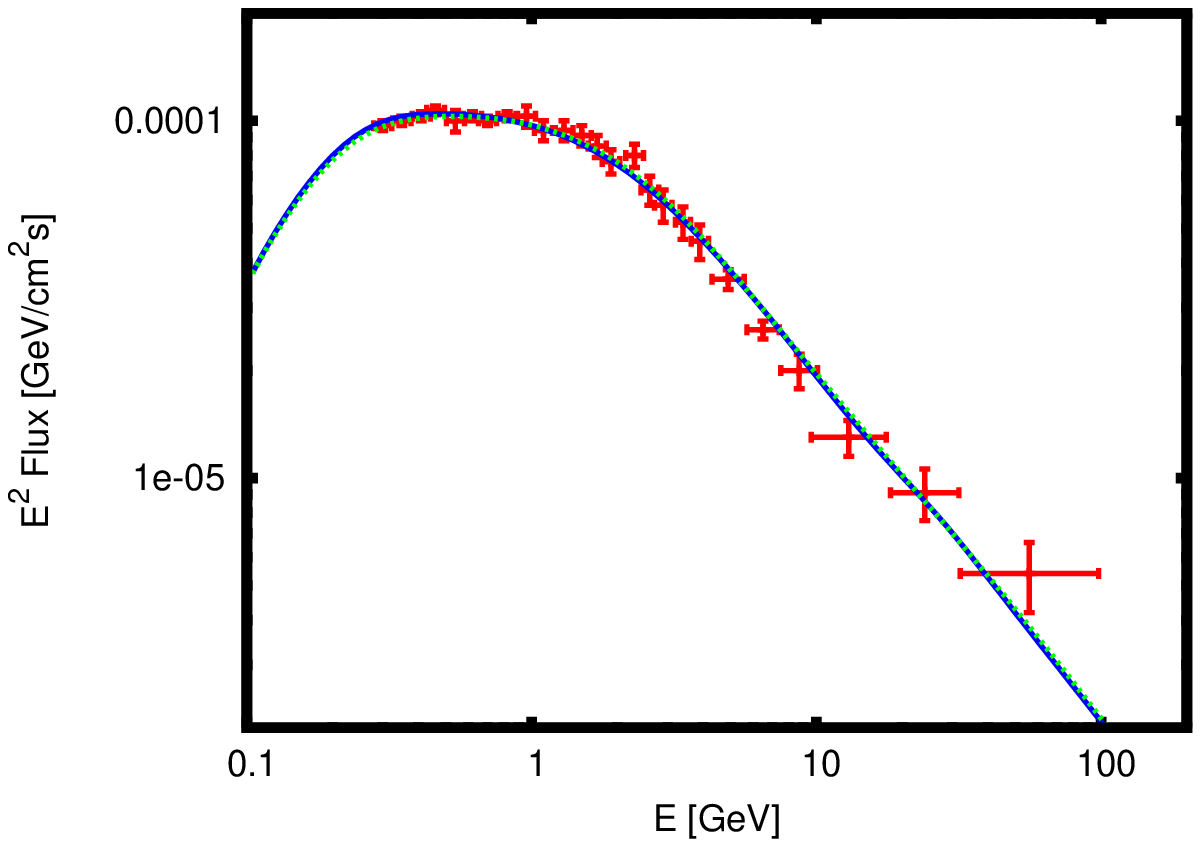}
\caption{Gamma-ray flux  $E^2\F(E)$  for a broken
  power-law with $\beta_{{\rm CR}}=-3$ for $p_{\rm CR}>10$\,GeV/c and
  $\beta_{{\rm CR}}=-2.4$ for $p_{\rm CR}<10$\,GeV/c 
  as calculated using the combination
of  QGSJET-II  and the ND model of Ref.~\cite{kamae07}
for $E_{\rm thr}=50$ GeV  (blue solid line) and 
based on the  Kamae parametrization alone (green dashed line);
All fluxes are normalized to coincide at 1\,GeV.
The gamma-ray spectrum derived in Ref.~\cite{NST} from Fermi-LAT data
is shown as points with error-bars.
\label{fig:gamma-b2}}
\end{figure}


As a further application of our new photon fragmentation functions, we 
consider the problem how the cosmic ray spectrum can be derived from
gamma-ray observations. The Fermi LAT collaboration has studied this
question in great detail and using a variety of methods~\cite{F1,F2,F3,other}: 
For instance, the diffuse Galactic gamma-ray emission was compared in 
Ref.~\cite{F1} to GALPROP models and an overall agreement of $\sim 15\%$ 
of data and models was found.
In Ref.~\cite{F2}, the analysis of the diffuse gamma-ray emissivity 
was constrained to well-defined segments of the Local and the Perseus 
arms, deriving thereby constraints on the cosmic ray density gradient.
Using observations in the mid-latitude region in the third quadrant,
the Fermi-LAT callaboration concluded in Ref.~\cite{F3} that the CR 
spectrum derived agrees with the locally measured one within $10\%$.

More recently, the authors of Ref.~\cite{NST} used Fermi-LAT observations of 
nearby molecular clouds to deduce the energy spectrum of Galactic sea CRs.
The photon flux deduced in Ref.~\cite{NST} is shown in Fig.~\ref{fig:gamma-b1}
as red error-bars  together with the photon spectrum (dashed green line) 
derived by us assuming a cosmic ray spectrum characterized by the
 slope $\beta_{{\rm CR}}=-2.85$, and using the combination of
  QGSJET-II results for $E_p>50$\,GeV with the parametrization
of Ref.~\cite{kamae07} for nondiffractive processes at lower energies.
The predicted photon flux agrees
well with the data at energies $E_\gamma\gsim 1$\,GeV.  The remaining 
discrepancy at lower energies is at most at 30\% level and may be partly 
related to uncertainties in the description of photon production at 
$E_{0}\lesssim 50$\,GeV~\cite{kamae07}. A more important reason for
the difference  at energies $E_\gamma\lsim 1$\,GeV may be the energy-dependence 
of the diffusion coefficient, which is expected to change 
at 2--3\,GeV~\cite{Ptuskin:2005ax,low2}: For instance, Ref.~\cite{low2}
found fitting Galactic synchrotron data that the diffusion coefficient 
reaches a minimum at 2--3\,GeV, increasing as $p^{-1}$ or even
faster at lower momenta $p$. To illustrate the latter point, we 
calculated  the diffuse gamma-ray flux for the case of
a broken power-law CR spectrum: with the  slope
 $\beta_{{\rm CR}}=-2.85$ for $p_{\rm CR}>3$\,GeV/c and
  $\beta_{{\rm CR}}=-1$ for $p_{\rm CR}<3$\,GeV/c.
  The result, plotted in Fig.~\ref{fig:gamma-b1}  as the
  solid blue line, matches well the observations in the whole energy
  range considered.
  
 Next we examine, if a spectral break at higher energies
  is also consistent with the Fermi-LAT data. In particular, 
 a broken power-law with high-energy slope 
   $\beta_{{\rm CR}}\simeq -3$, low-energy slope $\beta_{{\rm CR}}\simeq -1.9$, 
  and break-energy  at  $T=E_p-m= 9$\,GeV, or $E_p\simeq 10$\,GeV, 
  has been obtained in Ref.~\cite{NST} as best-fit to the Fermi-LAT data. 
  In Fig.~\ref{fig:gamma-b2},  we show  as blue solid line 
 the gamma-ray flux corresponding to a CR
 spectrum with the slope  $\beta_{{\rm CR}}=-3$ for $p_{\rm CR}>10$\,GeV/c
and  $\beta_{{\rm CR}}=-2.4$  for $p_{\rm CR}<10$\,GeV/c which agrees clearly 
with the data also well. Note that the sharper spectral break obtained in  
Ref.~\cite{NST}, with the slope $\beta_{{\rm CR}}=-1.9$ for $T<9$ GeV, 
is caused by the use of the power-law CR spectra with respect
to the kinetic energy in  Ref.~\cite{NST}, leading to a
substantial enhancement of the region of small $p_{\rm CR}$. 
To test the sensitivity of these results to the photon fragmentation
function used, we repeat the  calculation using the Kamae 
parametrization: We obtained a very similar photon spectral shape,
 as illustrated by the dashed green line 
in Fig.~\ref{fig:gamma-b2}, though with a 20\% higher flux.
 For an easier comparison of the shape,
we normalize all the fluxes to coincide at 1\,GeV.
 As we have shown above, the  difference between the various
 fragmentation model manifest themselves mainly in the absolute 
photon yield, not in the spectral shape.
  
Although the CR spectrum derived in Ref.~\cite{NST} is consistent with
the data, we consider it as less attractive:
First, this solution requires additionally to the break at 9\,GeV
another one around 200\,GeV, where the transition to the slope measured
by PAMELA should take place. Second, both break energies have no
obvious physical reason.  In particular, it is surprising that the
CR spectrum measured inside the Solar system differs substantially
from the one of Galactic ``sea'' CRs up to  200\,GeV. In contrast, the 
spectrum shown earlier in 
Fig.~\ref{fig:gamma-b1} has a single break at an energy where a change
of the diffusion coefficient is expected ~\cite{Ptuskin:2005ax,low2}
and a single  exponent 
$\beta_{{\rm CR}}=-2.85$ in agreement with PAMELA measurements.

\section{Conclusions}

We have reconsidered the problem of determining the average properties of 
Galactic CRs using gamma-ray observations of molecular clouds. The largely 
improved quality of the observational data requires a careful treatment 
of the photon fragmentation function. Comparing photon fragmentation 
functions calculated in different approaches at high energies, 
we have argued that a combination of the Kamae parametrisation and QGSJET-II 
provides the most reliable results. 
As our main result, we obtained that the spectral shape of CR protons 
as determined by PAMELA is consistent with 
the photon spectra from molecular clouds observed by Fermi-LAT down to 
energies $E\sim 1$--2\,GeV. The agreement is  improved further, if the
CR spectrum exhibits a break around 3\,GeV. This gives additional 
evidence for a change of the diffusion coefficient around 3\,GeV,
which was previously suggested on theoretical grounds~\cite{Ptuskin:2005ax} 
and supported by observations~\cite{low2}.

\acknowledgments

We would like to thank Andrii Neronov, Dima Semikoz, and Andrew Taylor for 
useful discussions, and Martin Pohl for comments on Ref.~\cite{huang}. 
S.O.\ acknowledges the support of Norsk Forskningsradet
within the program Romforskning.

\vskip0.5cm
{\it Note added:\/} After submission of this work, preprints by C.~Dermer
[arXiv:1206.2899] and by Malkov, Diamond, and Sagdeev [arXiv:1206.1384] 
appeared discussing also the connection of gamma-ray and CR spectra.

\appendix

\section{Using the fragmentation functions}
The photon fragmentation functions for proton-proton, proton-helium, and
helium-proton collisions have been
tabulated based on QGSJET-II simulations for a set of incident proton energies
$E_p$ between 10 and $10^8$ GeV. For given energies of the primary
 $E_{\rm p}$ and   secondary $E_{\rm s}$ particles 
 the inclusive spectrum $E_{\rm s}\:d\sigma (E_{\rm p},E_{\rm s})/dE_{\rm s}$
 is then obtained via an interpolation between the
 tabulated values. In the case of pp collisions, the interpolation between
the non-diffractive part of Kamae's parametrisation and QGSJET-II used in this 
work is provided. All procedures are available at 
\url{http://sourceforge.net/projects/ppfrag}.

\section{Comparison of the antiproton fragmentation function with data}
In addition to   photon production, we performed a comparison of model
predictions for antiproton spectra in proton-proton 
collisions with available data,
being motivated by the importance of those spectra in particular for 
dark matter searches.

\begin{figure*}[tbh]
\includegraphics[width=\textwidth]{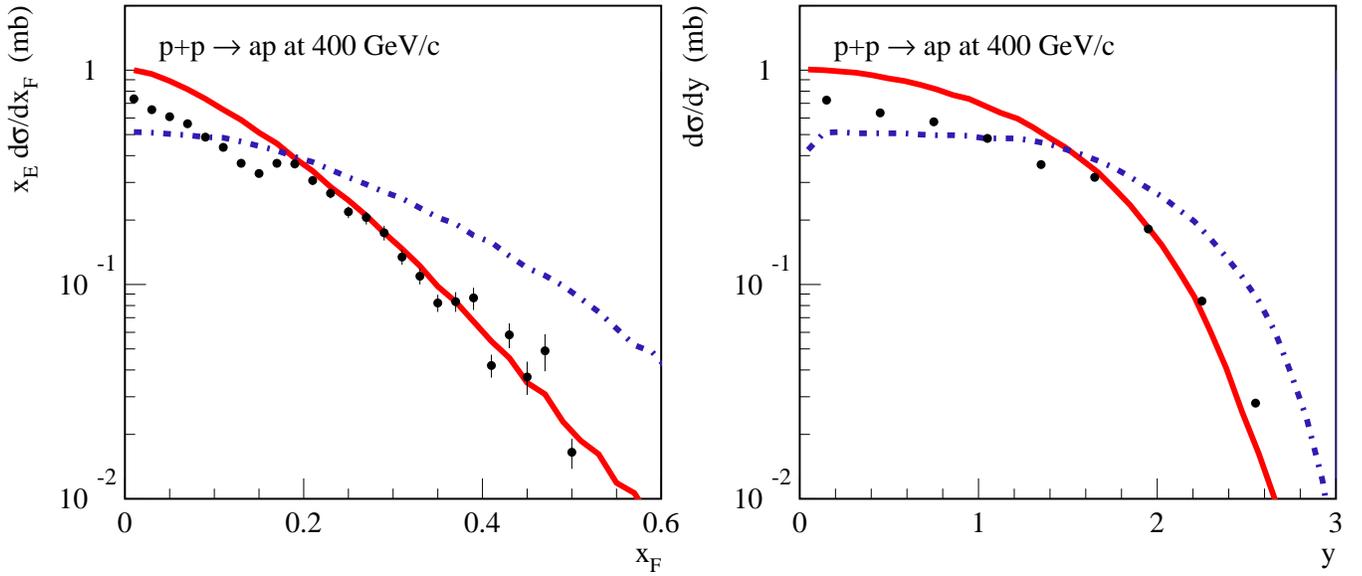}
\caption{Antiproton spectra in the c.m.\ frame
for proton-proton collisions at 400 GeV/c:   Feynman $x$ spectrum (left),
and rapidity distribution (right). Calculations with QGSJET-II-04
(red solid) and  SIBYLL~2.1 (blue dot-dashed)
 are shown together with experimental data from Ref.~\cite{ehs400}. 
\label{fig:ap400}}
\end{figure*}

\begin{figure*}[tbh]
\includegraphics[width=0.9\textwidth]{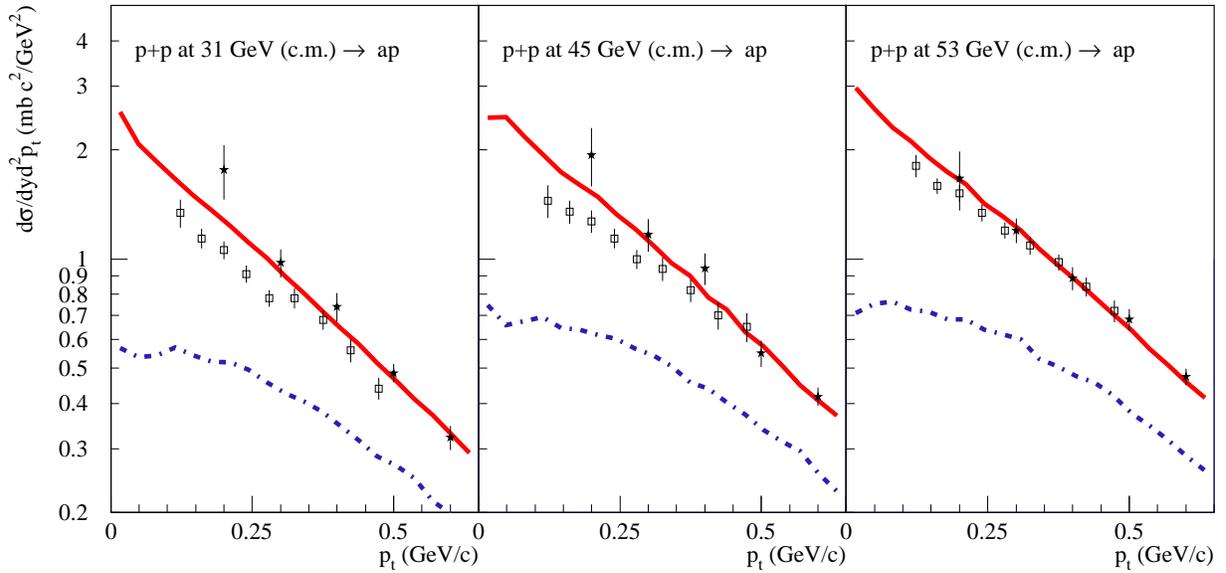}
\caption{Transverse momentrum spectra of antiprotons at $y=0$ 
in the c.m.\ frame at $\sqrt{s}=31$, 45, and 53\,GeV,
as calculated using  QGSJET-II-04
(red solid) and  SIBYLL~2.1 (blue dot-dashed) compared to 
 experimental data from Refs.~\cite{alper,guettler}.
\label{fig:ap31}}
\end{figure*}

As one can see from Figs.~\ref{fig:ap400} and \ref{fig:ap31}, pre-LHC data 
are somewhat uncertain concerning the antiproton yield in $pp$ collisions,
which partly explains the vast differences between various model
predictions.

\begin{figure}[t]
\includegraphics[width=0.45\textwidth]{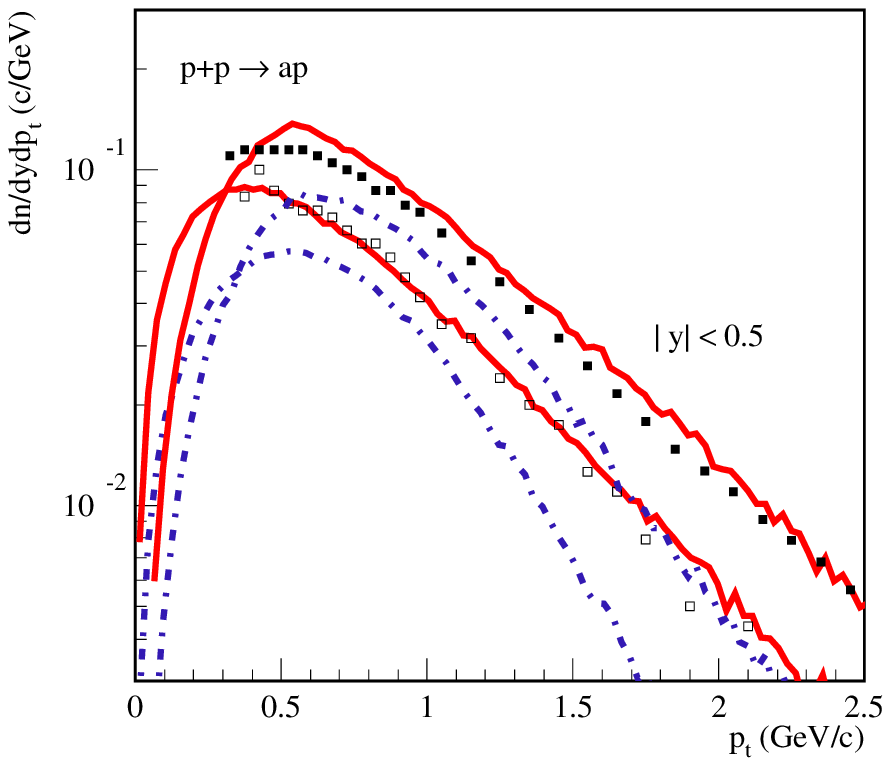}
\caption{Transverse momentrum spectra of antiprotons at central rapidities
($|y|<0.5$) 
in the c.m.\ frame at $\sqrt{s}=900$ GeV (open squares) and 7 TeV (full squares)
as measured by the ALICE Collaboration \cite{alice-ap} compared to 
 calculations with  QGSJET-II-04
(red solid) and  SIBYLL~2.1 (blue dot-dashed).
\label{fig:ap-lhc}}
\end{figure}

An important benchmark on the energy-dependence of the antiproton
production is provided by the recent data
obtained at 
LHC~\cite{alice-ap}, as illustrated in  Fig.~\ref{fig:ap-lhc}.
The antiproton yield 
in QGSJET-II-04 has been adjusted using these data.

Fragmentation functions for the production of antiprotons (plus antineutrons)
in proton-proton, proton-helium, and
helium-proton collisions are also available at 
\url{http://sourceforge.net/projects/ppfrag}.


\end{document}